\documentclass[a4paper]{jpconf} 
\usepackage{graphicx}
\usepackage{amssymb,amsmath} 
\def\lazz{\mathrel{\mathchoice {\vcenter{\offinterlineskip\halign{\hfil
$\displaystyle##$\hfil\cr<\cr\sim\cr}}}
{\vcenter{\offinterlineskip\halign{\hfil$\textstyle##$\hfil\cr<\cr\sim\cr}}}
{\vcenter{\offinterlineskip\halign{
\hfil$\scriptstyle##$\hfil\cr<\cr\sim\cr}}}
{\vcenter{\offinterlineskip\halign{\hfil$\scriptscriptstyle##
$\hfil\cr<\cr\sim\cr}}}}}

\def\pr{\prime}
\def\be{\begin{equation}}
\def\lan{\left\langle}
\def\ran{\right\rangle}
\def\ee{\end{equation}}
\def\barr{\begin{array}}
\def\earr{\end{array}}

\def\nn8{\\}
\def\l{\left}
\def\r{\right}
\def\dis{\displaystyle}
\def\ed{\end{document}}

\def\cod{{\cal O}^\dagger}
\def\co{{\cal O}}

\def\wtM{{\widetilde {M}}}

\begin{document}
\title{Random Matrix Theory with U(N) Racah Algebra for Transition Strengths}

\author{V.K.B. Kota}

\address{Physical Research Laboratory, Ahmedabad 380 009, India}

\ead{vkbkota@prl.res.in}

\begin{abstract}

For finite quantum many-particle systems, a given system, induced by a
transition operator, makes transitions from its states to the states of the same
system or to those of another system. Examples are electromagnetic transitions
(then the initial and final systems are same), nuclear beta and double beta
decay (then the initial and final systems are different), particle addition to
or removal from a given system and so on. Working towards developing a complete
statistical theory for transition strength densities (transition strengths
multiplied by the density  of states at the initial and final energies), we have
started a program to  derive formulas for the lower order bivariate moments of
the strength densities generated by a variety of transition operators. In this
paper results are presented for a transition operator that removes $k_0$ number
of particle by considering $m$ spinless fermions in $N$ single particle states.
The Hamiltonian  that is $k$-body is represented by EGUE($k$) [embedded Gaussian
unitary ensemble of $k$-body interactions] and similarly the transition operator
by an appropriate independent EGUE. Employing the embedding $U(N)$  algebra,
finite-$N$ formulas for moments  up to order four are derived and they show that
in general the smoothed (with respect to energy) bivariate transition strength 
densities take bivariate Gaussian form. Extension of these results to particle
addition operator and  beta decay type operators are discussed.

\end{abstract}

\section{Introduction}

Let us begin with the statement from the preface to the proceedings of the
meeting held in 2006 on "Applications of Random Matrices in Physics":
\cite{quote} {\it Random matrices are widely and successfully used in physics
for almost 60-70 years, beginning with the works of Wigner and Dyson. Initially
proposed to describe statistics of levels in complex nuclei, the Random Matrix
Theory has grown far beyond nuclear physics, and also far beyond just level
statistics. It is constantly developing into new areas of physics and
mathematics, and now constitutes a part of the general culture and curriculum of
a theoretical physicist.} Besides  applications in all branches of quantum
physics, RMT is being used in disciplines such as Econophysics, Wireless
communication, information theory, multivariate statistics, number theory,
neural and biological networks and so on. The focus in this article is on the
frontier topic of statistical properties of isolated finite many-particle
quantum systems \cite{SenRMP}.  Examples for these systems are atoms, atomic
nuclei, mesoscopic systems (quantum dots, small metallic grains), interacting
spin systems modeling quantum computing core, ultra-cold atoms and so on. A
route to investigate statistical properties is to employ the classical GOE or
GUE or GSE random matrix ensembles with various deformations. For these, as
Wigner states: {\it  the assumption is that the Hamiltonian which governs the
behavior of a complicated system is a random real symmetric or complex Hermitian
or Quaternion real matrix, with no special properties except for its symmetric
or Hermitian or Quaternion real nature.}  However, for most of the isolated
finite many-particle quantum systems, their constituents predominantly interact
via two-particle interactions and the classical random matrix ensembles are too
unspecific to account for this most important feature.  One refinement which
retains the basic stochastic  approach but allows for this feature consists in
the use of embedded random matrix ensembles
\cite{Ko-01,Br-81,BW-JPA,PW,Go-12,kota}. 

Therefore, it is more appropriate to represent an isolated  finite interacting
quantum system, say with $m$ particles (fermions or bosons) in $N$ single
particle (sp) states by random matrix models generated by random $k$-body (note
that $k < m$ and most often we have $k=2$)  interactions and propagate the
information in the interaction to many particle spaces. Thus we have random
interaction matrix models for $m$-particle systems. In the simplest version, the
$k$-particle Hamiltonian ($H$) of a spinless fermion (or boson) system is
represented by GOE/GUE/GSE and then the $m$ particle $H$ matrix is generated
using the $m$-particle Hilbert space geometry. The key element here is  the
recognition that the $U(N)$ Lie algebra  transports the information in the
two-particle spaces to many-particle spaces.  As a GOE/GUE/GSE random matrix
ensemble in two-particle spaces is embedded  in the $m$-particle $H$ matrix,
these ensembles are generically called embedded ensembles (EE). With GUE
embedding we have EGUE and in this paper EGUE is used throughout. For EE, a
general formulation for deriving analytical  results is to use the Wigner-Racah
algebra of the embedding Lie algebra \cite{kota}. The focus in the present paper
is on transition strengths and they are not yet studied in any detail using EE
\cite{kota}. Also, as emphasized in \cite{Br-81,Widen}, there are many open
questions in the random matrix theory for transition strengths in finite
interacting quantum many-particle systems.

For finite quantum many-particle systems, induced by a transition operator, a
given system makes transitions from its states to the states of the same system
or to the states of another system. Examples are electromagnetic transitions
(then the initial and final systems are same), nuclear beta and double beta
decay (then the initial and final systems are different), particle addition to
or removal from a given system and so on. Given a state with energy $E_i$ and
say it is connected to a  state $E_f$ by a transition operator $\co$, then the
transition strengths are $\l|\lan E_f \mid \co \mid E_i\ran\r|^2$ and these will
determine for example the life time  of a state. It is important to recognize
that the transition strengths probe the  structure of the eigenfunctions of a
quantum many-body system and thus they are very important from the point of view
of experiments  probing the structure of a system. Also, they are needed in many
applications (for example, beta decay transition strengths are essential for
nucleosynthesis studies). In the statistical theories, it is more useful to deal
with the corresponding transition strength density (this will take into account
degeneracies in the eigenvalues) defined by
\be
I_\co(E_i,E_f) = I(E_f)\,\l|\lan E_f \mid \co \mid E_i\ran\r|^2\,I(E_i) \;.
\label{strn-eq1}
\ee
In Eq. (\ref{strn-eq1}), $I(E)$ are state densities normalized to the dimension
of the $m$ particle spaces. Note that $E_i$ and $E_f$ belong to the same $m$
particle system or different systems depending on the nature of $\co$, the 
transition operator.

Working towards developing a complete statistical theory for transition strength
densities (transition strengths multiplied by the density  of states at the
initial and final energies) for isolated finite many-particle quantum systems,
we have started a program to  derive formulas for the lower order bivariate
moments of the strength densities generated by a variety of transition
operators.   In general, the Hamiltonian may have many symmetries with the
fermions (or bosons) carrying other degrees of freedom such as spin, orbital
angular momentum, isospin and so on. Also we may have in the system different
types of fermions (or bosons) and for example in atomic nuclei we have protons
and neutrons. In addition, a transition operator may preserve particle number
and other quantum numbers or it may change them. Among all these various
situations, in our program we have considered five different  systems: (i) a
system of $m$ spinless fermions and a transition operator that preserves  the
particle number;  (ii) a system of $m$ spinless fermions and a transition
operator that  removes say $k_0$ number of particles from the $m$ fermion
system;  (iii) same as (ii) but for particles addition operator;  (iv) a system
with two types of spinless fermions with the transition operator changing $k_0$
number of particles of  one type to $k_0$ number of other particles as in
nuclear beta and double beta decay; (v) same as (i)-(iii) but for spinless boson
systems. In \cite{group30}, results are presented for (i). In the present paper
results are presented for (ii) and their extensions to (iii) and (iv) are
briefly discussed; Ref. \cite{Ma-arx} gives results from a first study for (iv).
Let us add that (ii) and (iii) are important for example in nuclear physics as
one and two particle removal and addition to a nucleus are important
experimental probes of the structure of  the atomic nucleus; see for example
\cite{Sch-12,Fr-07}. Thus, the results in Sections 3-5 have applications in
nuclear physics. Now we will give a preview.

Section 2 gives some basic results for EGUE($k$) for spinless fermion systems as
derived in \cite{Ko-05}. Using these results,
formulas for the lower order bivariate moments of the transition strength
densities for the situation (ii) above are derived and they are presented in
detail in Section 3. Using these, results in the asymptotic limit are derived
and they are presented in Section 4. Extension of the results in Section 3
to the situations (iii) and (iv) above are briefly discussed in Section 5.
Finally, in Section 6 gives conclusions and future outlook.

\section{Basic EGUE($k$) results for a spinless fermion system}

Let us consider $m$ spinless fermions in $N$ degenerate sp states with the 
Hamiltonian $\hat{H}$ a $k$-body operator,
\be
\hat{H} = \sum_{i,j} V_{ij}(k)\;A^\dagger_i(k)\,A_j(k)\;,\;\;\;V_{ij}(k)
=\lan k,i \mid \hat{H} \mid k,j\ran\;.
\label{strn-eq2}
\ee
Here $A^\dagger_i(k)$ is a $k$ particle (normalized) creation operator and
$A_i(k)$ is the corresponding annihilation operator (a hermitian conjugate). 
Also, $i$ and $j$ are $k$-particle indices. The $k$ and $m$ particle
space dimensions are $\binom{N}{k}$ and $\binom{N}{m}$ respectively. 
Representing the 
$V$ matrix, defined by the matrix elements $V_{ij}$, by GUE we have EGUE($k$)
for the $H$ matrix in $m$-particle spaces. Note that, for $V$ a GUE,
the real and imaginary parts of $V_{ij}$
are independent zero centered Gaussian random variables with variance 
satisfying,
\be
\overline{V_{ab}(k)\, V_{cd}(k)} = V^2_H\,\delta_{ad}\delta_{bc}\;.
\label{strn-eq3}
\ee
Here the 'over-line' indicates ensemble average. From now on we will drop the hat
over $H$ and denote when needed  $H$ by $H(k)$. In physical
systems in general $k=2$ but in some systems such as atomic nuclei and BEC
it is possible to have $k=3$ and even $k=4$ \cite{Zel3,Fu-13,BEC}.

The  $U(N)$ algebra that generates the embedding, as shown in \cite{Ko-05},
gives formulas for the lower order moments of the one-point function, the
eigenvalue density $I(E)=\overline{\lan\lan \delta(H-E)\ran\ran}$ and also for
the two-point function in the eigenvalues.  Used here is the $U(N)$
tensorial decomposition of the  $H(k)$ operator giving $\nu=0,1,\ldots,k$
irreducible parts $B^{\nu , \omega_\nu}(k)$ and  then,
\be
H(k) = \dis\sum^{k}_{\nu =0;\omega_\nu \in \nu} 
W_{\nu , \omega_\nu}(k)\; B^{\nu , \omega_\nu}(k) \;.
\label{strn-eq5}
\ee
Note that $\omega_\nu$ are labels of the irreducible representations (irreps)
of the subalgebras of $U(N)$ and their explicit structure will not play any 
role in the present work. With the GUE($k$) representation for the $H(k)$ 
operator, the expansion coefficients W's will be independent zero centered 
Gaussian random variables with
\be
\overline{W_ {\nu_1 , \omega_{\nu_1}}(k)\;W_ {\nu_2 , \omega_{\nu_2}}(k)}
= V^2_H\;\delta_{\nu_1 , \nu_2} \delta_{\omega_{\nu_1} \omega_{\nu_2}}\;.
\label{strn-eq6}
\ee
For deriving formulas for the various moments, the first step is to apply the
Wigner-Eckart theorem for the matrix elements of $B^{\nu , \omega_\nu} (k)$.
Given the $m$-fermion states $\l.\l|f_m v_i\r.\ran$, we have with respect to the
$U(N)$ algebra, $f_m=\{1^m\}$, the antisymmetric irrep in
Young tableaux notation and $v_i$ are additional labels. Note that the $\nu$
label used for $B$'s corresponds to the Young tableaux $\{2^\nu 1^{N-2\nu}\}$.
Now, Wigner-Eckart theorem for $U(N) \supset G$ (here $G$ is some subalgebra of
$U(N)$ giving $v$ and $\omega$ labels) gives
\be
\lan f_m v_f \mid B^{\nu , \omega_\nu}(k) \mid f_m v_i\ran = \lan f_m 
\mid\mid B^{\nu}(k) \mid\mid f_m\ran\;C^{\nu , \omega_\nu}_{f_m v_f\,,\;
\overline{f_m} \overline{v_i}} \;.
\label{W-E}
\ee
Here, $\lan --|| -- || --\ran$ is the reduced matrix element and 
$C^{----}_{----}$ is a $U(N) \supset G$ Clebsch-Gordan (C-G) coefficient 
[note that we are
not making a distinction between $U(N)$ and $SU(N)$]. Also, if $\l.\l|f_r
v\r.\ran$ represents a state for $r$ number of fermions, then $\l.\l|
\overline{f_r} \overline{v}\r.\ran$ represents the corresponding state for
$r$ number of holes or $N-r$ number of fermions (see
\cite{Ko-05} for details). In Young tableaux notation $\overline{f_m} =
\{1^{N-m}\}$. Definition of  $B^{\nu , \omega_\nu}(k)$  and the $U(N)$
Wigner-Racah algebra will give,
\be
\barr{l}
\l|\lan f_m \mid\mid B^{\nu}(k) \mid\mid f_m\ran\r|^2 = 
\Lambda^{\nu}(N,m,m-k)\;,
\;\\
\\
\Lambda^{\mu}(N^\pr,m^\pr,r) = \dis\binom{m^\pr-\mu}{r}\,\dis\binom{
N^\pr-m^\pr+r-\mu}{r}\;.
\earr \label{reduced}
\ee
The $\Lambda^{\nu}(N,m,k)$ are nothing but, apart from a $N$ and $m$
dependent factor, a $U(N)$ Racah coefficient \cite{Ko-05}. This and the various
properties of the $U(N)$ Wigner and Racah coefficients give two formulas for the
ensemble  average of a product any two $m$ particle matrix elements of $H$,
\be
\barr{l}
\overline{\lan f_m v_1 \mid H(k) \mid f_m v_2\ran\, \lan f_m v_3 \mid H(k) 
\mid f_m v_4\ran} \\
= V^2_H\;\dis\sum_{\nu=0;\omega_\nu}^{k}\Lambda^{\nu}(N,m,m-k)\; C^{\nu , 
\omega_\nu}_{
f_m v_1\,,\;\overline{f_m} \overline{v_2}}\, C^{\nu , \omega_\nu}_{
f_m v_3\,,\;\overline{f_m} \overline{v_4}}\;,
\earr \label{matrix-ele-a}
\ee
and also
\be
\barr{l}
\overline{\lan f_m v_1 \mid H(k) \mid f_m v_2\ran\, \lan f_m v_3 \mid H(k) 
\mid f_m v_4\ran} \\
= V^2_H\;\dis\sum_{\nu=0;\omega_\nu}^{m-k}\Lambda^{\nu}(N,m,k)\; 
C^{\nu , \omega_\nu}_{
f_m v_1\,,\;\overline{f_m} \overline{v_4}}\, C^{\nu , \omega_\nu}_{
f_m v_3\,,\;\overline{f_m} \overline{v_2}}
\earr \label{matrix-ele-b}
\ee
Eq. (\ref{matrix-ele-b}) follows by applying a Racah transform to the
product of the two C-G coefficients appearing in Eq. (\ref{matrix-ele-a}). 
Let us mention two important properties of the $U(N)$ C-G coefficients that 
are quite useful,
\be
\dis\sum_{v_i} C^{\nu , \omega_\nu}_{f_m v_i\,,\,\overline{f_m} 
\overline{v_i}} =\dis\sqrt{\binom{N}{m}}\;\delta_{\nu , 0}\;,\;\;\;
C^{0,0}_{f_m v_i\,,\;\overline{f_m}
\overline{v_j}}={\binom{N}{m}}^{-1/2}\;\delta_{v_i\,,v_j}\;.
\label{sum00}
\ee 
From now on we will use the symbol $f_m$ only in the C-G coefficients, Racah 
coefficients and the  reduced matrix elements. However, for the matrix elements
of an operator we will use $m$ implying totally antisymmetric state for 
fermions. An important by-product of Eqs. (\ref{matrix-ele-b}) and 
(\ref{sum00}) is
\be
\dis\sum_{v_j}\,\overline{\lan m v_i \mid H(k) \mid m v_j\ran\, 
\lan m v_j \mid H(k) 
\mid m v_k\ran} = \overline{\lan [H(k)]^2\ran^m}\;\delta_{v_i , v_k} 
\label{sum-hh}
\ee
and we will use this in Section 3.  

Starting with Eq. (\ref{strn-eq5}) and using Eqs. (\ref{strn-eq6}),
(\ref{matrix-ele-b}) and (\ref{sum00}) will immediately give the formula,  
\be
\overline{\lan [H(k)]^2\ran^m} = {\binom{N}{m}}^{-1}\;\dis\sum_{v_i}\,
\overline{\lan m v_i \mid [H(k)]^2 \mid m v_i\ran} = V^2_H \;\Lambda^0(N,m,k)\;.
\label{strn-eq7a}
\ee
Similarly, for $\overline{\lan H^4\ran^m}$ first the ensemble average is
decomposed into 3 terms as, 
\be
\barr{l}
\overline{\lan\lan [H(k)]^4\ran\ran^m} = \dis\sum_{v_i}\,
\overline{\lan m v_i \mid [H(k)]^4 \mid m v_i\ran} \\
\\
= \dis\sum_{v_i, v_j, v_p , v_l} \l[\overline{\lan m v_i \mid H(k) \mid m 
v_j\ran\,\lan m v_j \mid H(k) \mid m v_p\ran}\;\; \overline{\lan m v_p 
\mid H(k) \mid m 
v_l\ran \lan m v_l \mid H(k) \mid m v_i\ran}\r. \\
\\
+ \overline{\lan m v_i \mid H(k) \mid m v_j\ran\,
\lan m v_l \mid H(k) \mid m v_i\ran}\;\; \overline{\lan m v_j \mid H(k) \mid m 
v_p\ran \lan m v_p \mid H(k) \mid m v_l\ran} \\
\\
+ \l. \overline{\lan m v_i \mid H(k) \mid m v_j\ran\,
\lan m v_p \mid H(k) \mid m v_l\ran}\;\; \overline{\lan m v_j \mid H(k) \mid m 
v_p\ran \lan m v_l \mid H(k) \mid m v_i\ran} \r] \;.
\earr \label{strn-eq7ab}
\ee
Note that the trace $\lan\lan H^4\ran\ran^m = \binom{N}{m} \lan H^4\ran^m$.
It is easy to see that the first two terms simplify to give $2[\overline{\lan
H^2\ran^m}]^2$ and the third term is simplified by applying Eq.
(\ref{matrix-ele-a}) to the first ensemble average and Eq. (\ref{matrix-ele-b})
to the second ensemble average. Then, the final result is
\be
\overline{\lan [H(k)]^4\ran^m} = 2 \l[ \overline{\lan H^2\ran^m}\r]^2 
+ V^4_H \;{\binom{N}{m}}^{-1}\; \dis\sum_{\nu=0}^{min(k,m-k)}\,
\Lambda^{\nu}(N,m,k)\, \Lambda^{\nu}(N,m,m-k)\,d(N:\nu)\;;
\label{strn-eq7b}
\ee
\be
d(N:\nu) = {\dis\binom{N}{\nu}}^2 -{\dis\binom{N}{\nu -1}}^2\;.
\label{eq-dnu}
\ee
Now, we will derive the formulas for the moments of the transition strength 
densities generated by a transition operator $\co$ that removes $k_0$ number 
of particles from a $m$-particle system.

\section{Lower-order moments of transition strength densities:
results for particle removal operators}

Particle removal (or addition) operators are of great interest in nuclear
physics. For example one particle (proton or neutron) removal from a target 
nucleus gives information about the single particle levels in the target and
similarly, two-particle removal gives information about pairing force.  
Let us begin with a particle removal operator $\co$ and say it removes $k_0$ 
number of particles when acting on a $m$ fermion state. Then the general form  
of $\co$ is,
\be
\co = \dis\sum_{\alpha_0} V_{\alpha_0}\;A_{\alpha_0}(k_0) \;.
\label{snt-1}
\ee
Here, $A_{\alpha_0}(k_0)$ is a $k_0$ particle annihilation operator and
$\alpha_0$ are indices for a $k_0$ particle state. Note that $A_{\alpha_0}(k_0)$
transforms as $\{\overline{f_{k_0}}\} = \{1^{N-k_0}\}$ with respect to $U(N)$
and $A^\dagger_{\alpha_0}(k_0)$  transforms as $\{f_{k_0}\}$. It important to
recognize that the $\co$ matrices will be rectangular matrices connecting $m$
particle states to $m-k_0$ particle states. In the defining space, the matrix
will be a $1 \times d_0$ matrix with matrix elements given by $V_{\alpha_0}$.
Note that $\alpha_0$ takes $d_0$ values and $d_0=\binom{N}{k_0}$.
We will represent $\co$ by EGUE implying that the defining space matrix elements
$V_{\alpha_0}$ are zero centered independent Gaussian random variables [also
they are independent of the $V_{ij}(k)$ variables in Eq. (\ref{strn-eq2}) 
and therefore also independent of the $W$ variables in Eq. (\ref{strn-eq5})]
with variance satisfying 
\be
\overline{V_{\alpha} V^{\dagger}_\beta} =  V^2_\co\;\delta_{\alpha
\beta} \;.
\label{snt-2}
\ee
In many particle spaces the $\co$ matrix will be a $d_1 \times d_2$ matrix
connecting $d_1=\binom{N}{m}$ number of $m$-particle states to
$d_2=\binom{N}{m-k_0}$ number of $(m-k_0)$-particle states. Using Eqs.
(\ref{snt-1}) and (\ref{snt-2}), we have 
\be
\overline{\lan \cod \co\ran^m} = V^2_\co\;\binom{m}{k_0}\;,\;\;\;
\overline{\lan \co \cod\ran^m} = V^2_\co\;\binom{N-m}{k_0}\;.
\label{snt-3}
\ee
Similarly, Eq. (\ref{sum-hh}) gives the relations,
\be
\overline{\lan \cod \co H^p\ran^m} = \overline{\lan \cod \co\ran^m}\;\; 
\overline{\lan H^p\ran^m}\;,\;\;\;
\overline{\lan \cod H^p \co\ran^m} = \overline{\lan \cod \co\ran^m}
\;\; \overline{\lan H^p\ran^{m-k_0}}\;.
\label{snt-4}
\ee
Another useful result follows by introducing complete set of states between the
$\cod$ and $\co$ operators in Eq. (\ref{snt-3}) and applying the Wigner-Eckart
theorem,
\be
\lan m \mid\mid A^{\dagger}(k_0) \mid\mid m-k_0\ran \lan m-k_0 \mid\mid
A(k_0) \mid\mid m\ran = \binom{N-k_0}{m-k_0} \;.
\label{snt5}
\ee
Following the procedure used in Section 2, it is possible to derive formulas for
the lower order moments of the transition strength densities generated by $\co$ 
defined by Eq. (\ref{snt-1}). The bivariate moments are defined by
\be
M_{PQ} = \overline{\lan \cod H^Q \co H^P\ran^m}\;.
\label{snt-6}
\ee
and we will consider the moments $P+Q=2$ and $4$ (the $P+Q=3$ moments are zero
as we are using independent EGUE representations for $\co$ and $H$ matrices).

Firstly, Eqs. (\ref{snt-4}) gives,
\be
\barr{l}
M_{20} = \overline{\lan \cod \co \ran^m}\;\;\overline{\lan H^2\ran^m}\;,\;\;\;
M_{02} = \overline{\lan \cod \co \ran^m}\;\;\overline{\lan H^2\ran^{m-k_0}}\;, 
\\
M_{40} = \overline{\lan \cod \co \ran^m}\;\;\overline{\lan H^4\ran^m}\;,\;\;\;
M_{04} = \overline{\lan \cod \co \ran^m}\;\;\overline{\lan H^4\ran^{m-k_0}}\;.
\earr \label{snt-7}
\ee
Now, Eq. (\ref{snt-3}) along with Eqs. (\ref{strn-eq7a}) and (\ref{strn-eq7b}) 
will give the formulas for $M_{20}$, $M_{02}$, $M_{40}$ and $M_{04}$.
Formula for the first non-trivial moment $M_{11}=\overline{\lan \cod H \co
H\ran^m}$ is derived by introducing complete set of states between $\cod$ and
$H$, $H$ and $\co$ and $\co$ and $H$ in the trace giving, 
\be 
\barr{l} 
\binom{N}{m} \;M_{11}(m) = \binom{N}{m}\;\overline{\lan \cod H \co H\ran^m} =  
\dis\sum_{v_1, v_2, v_3, v_4} \overline{\lan m, v_1 \mid \cod \mid m-k_0, 
v_2\ran \lan m-k_0, v_3 \mid \co \mid  m,
v_4\ran}\\ 
\times\;\overline{\lan m-k_0, v_2 \mid H \mid m-k_0, v_3\ran \lan  m,
v_4 \mid H \mid m, v_1\ran}\;.  
\earr \label{snt-8a} 
\ee 
Using Eq. (\ref{snt-1}) and applying Eq. (\ref{snt-2}) along with
Eqs. (\ref{strn-eq5}) - (\ref{reduced}) and the Wigner-Eckart theorem  
will give, 
\be 
\barr{l} 
M_{11}(m) = V_{\co}^2 V_H^2\; {\binom{N}{m}}^{-1}\; \binom{N-k_0}{m-k_0}\;
\dis\sum_{\nu=0}^k \l[\Lambda^{\nu}(N,m-k_0,m-k_0-k)\; \Lambda^{\nu}
(N,m,m-k)\r]^{1/2} \\
\times\; \dis\sum_{v_1, v_2, v_3, v_4;\,\alpha;\,\omega_\nu} 
C^{f_{k_0} , \alpha}_{f_m v_1\,,\;\overline{f_{m-k_0}} \overline{v_2}}\,
C^{\overline{f_{k_0}} , \overline{\alpha}}_{f_{m-k_0} v_3\,,\;\overline{f_{m}} 
\overline{v_4}}\,
C^{\nu , \omega_{\nu}}_{f_{m-k_0} v_2\,,\;\overline{f_{m-k_0}} 
\overline{v_3}}\,
C^{\nu , \omega_{\nu}}_{f_m v_4\,,\;\overline{f_m} 
\overline{v_1}}\;.
\earr 
\label{snt-8b} 
\ee 
Simplifying the four C-G coefficients will give finally,
\be
\barr{l}
M_{11}(m) = V_{\co}^2 V_H^2\; {\binom{N}{m}}^{-1}\; \binom{N-k_0}{m-k_0}\;
\dis\sum_{\nu=0}^k Z_{11}(N, m, k_0, k, \nu)\;;\\
Z_{11}(N, m, k_0, k, \nu) = \l[\binom{N}{k_0}\,d(N : \nu)\,
\Lambda^{\nu}(N,m,m-k)\,\Lambda^{\nu}(N,m-k_0,m-k_0-k)\r]^{1/2} \\
\\
\times (-1)^{\phi(f_{m}, \overline{f_{m-k_0}}, f_{k_0}) + \phi(f_{m-k_0}, 
\overline{f_{m-k_0}}, \nu)}\; U(f_{m}\,\overline{f_{m-k_0}}\,
f_{m}\,f_{m-k_0}\,;\,f_{k_0}\,\nu)\;.
\earr \label{snt-8}
\ee
Here $\phi$ is a phase factor and it is a function of the $U(N)$ irreps. It is
shown elsewhere (V.K.B. Kota and Manan Vyas, in preparation) that $
(-1)^{\phi(f_{m}, \overline{f_{m-k_0}}, f_{k_0}) + \phi(f_{m-k_0}, 
\overline{f_{m-k_0}}, \nu)}\; U(---)$ will be positive where $U(--)$ is a $U(N)$
$U$-coefficient. Therefore we need only $U^2$ and the formula for this is given
by \cite{He-75},
\be
\l[U(f_m,\, \overline{f_p},\, f_m,\, f_p\, ;\,f_{m-p}\,\nu)\r]^2 = 
\dis\frac{{\binom{N+1}{\nu}}^2 \binom{m-\nu}{p-\nu} \binom{N-\nu -p}{m-p}\;
(N-2\nu +1)}{{\binom{N-m+p}{p}}^2 \binom{N}{m-p}\;(N+1)}\;.
\label{snt-9}
\ee

Turning to the fourth order moments, we need $M_{13}$, $M_{31}$ and $M_{22}$.
As $\cod \neq \co$, here $M_{13} \neq M_{31}$ [similarly $M_{40} \neq M_{04}$ 
and $M_{20} \neq M_{02}$ as seen from Eq. (\ref{snt-7})]. Following the 
procedure used for deriving the formula for $M_{11}(m)$, we have for 
$M_{31}(m)$
\be
\barr{l}
M_{31}(m) = \overline{\lan \cod H \co H^3 \ran^{m}} = 
2\;\overline{\lan H^2\ran^{m}}\;M_{11}(m) \\
\\
+\; V^2_\co\;V^2_H\;\binom{N}{m}^{-1} \;\dis\sum_{v_1, v_2, 
v_3, v_4, v_5, v_6: \alpha , \nu_1, \omega_{\nu_1}, \nu_2, \omega_{\nu_2}}  \\
\\
\lan m, v_1 \mid A^{\dagger}_{\alpha}(k_0) \mid m-k_0, v_2 \ran \lan m-k_0, 
v_3 \mid A_{\alpha}(k_0)\mid m, v_4\ran \\
\\
\times \lan m-k_0, v_2 \mid B^{\nu_1, \omega_{\nu_1}}(k) \mid m-k_0, 
v_3\ran \lan m, v_5 \mid B^{\nu_1, \omega_{\nu_1}}(k) \mid m, v_6\ran \\
\\
\times\; \lan m, v_4 \mid B^{\nu_2, \omega_{\nu_2}}(k) \mid m, 
v_5\ran \lan m, v_6 \mid B^{\nu_2, \omega_{\nu_2}}(k) \mid m, v_1\ran\;.
\earr \label{snt-10}
\ee
Now, applying the Wigner-Eckart theorem, using the results in Section 2 and
simplifying the resulting C-G coefficients will give,
\be
\barr{l}
M_{31}(m) = \overline{\lan \cod H \co H^3 \ran^{m}} = 
2\;\overline{\lan H^2\ran^{m}}\;M_{11}(m) \\
\\
+ V^2_\co \, V^2_H\;\binom{N}{m}^{-1}\; \binom{N-k_0}{m-k_0}\; 
\dis\sum_{\nu=0}^{min(k,m-k)} \Lambda^{\nu}(N,m,k)\;
Z_{11}(N, m, k_0, k, \nu)\;.
\earr \label{snt-11}
\ee
The function $Z_{11}$ is defined in Eq. (\ref{snt-8}). Following the same 
procedure as above, the formula for $M_{13}$ is,
\be
\barr{l}
M_{13}(m) = \overline{\lan \cod H^3 \co H \ran^{m}} = 
2\;\overline{\lan H^2\ran^{m-k_0}}\;M_{11}(m) \\
\\
+ V^2_\co\,V^2_H\;  \binom{N}{m}^{-1}
\binom{N-k_0}{m-k_0}\;  \dis\sum_{\nu=0}^{min(k,m-k_0-k)} 
\Lambda^{\nu}(N, m-k_0, k)\;Z_{11}(N, m, k_0, k, \nu)\;.
\earr \label{snt-12}
\ee   
Derivation of the formula for $M_{22}$ is more involved. Leaving details to a
long paper under preparation, the final result (with $\rho$ a multiplicity
label) is
\be
\barr{l}
M_{22}(m) = \overline{\lan \cod H^2 \co H^2\ran^m} =
\overline{\lan \cod \co\ran^{m}}\;\overline{\lan H^2\ran^{m}}\;
\overline{\lan H^2\ran^{m-k_0}} \\
\\
+ V^2_\co \;V^2_H \,\l\{\binom{N}{m} \binom{N}{k_0}\r\}^{-1}
\binom{N-k_0}{m-k_0}\;\l\{\dis\sum_{\nu=0}^k Z_{11}(N, m, k_0, k, \nu)\r\}^2 \\
\\
+ V^2_\co \;V^2_H\, \binom{N}{m}^{-1} \binom{N-k_0}{m-k_0}\;
\dis\sum_{\nu_1=0}^{k}
\dis\sum_{\nu_2=0}^{k} \dis\sum_{\nu=0}^{2k} 
\dis\sqrt{\binom{N}{k_0}\;d(N : \nu)} \\
\\
\times \dis\sum_\rho \lan m \mid\mid \l[B^{\nu_1}(k) B^{\nu_2}(k)\r]^{
\nu:\rho} \mid\mid m\ran\;\lan m-k_0 \mid\mid \l[B^{\nu_1}(k) 
B^{\nu_2}(k)\r]^{\nu:\rho} \mid\mid m-k_0\ran \\
\\
\times\; (-1)^{\phi(f_{m},\overline{f_{m-k_0}}, k_0) + \phi(f_{m-k_0},
\overline{f_{m-k_0}}, \nu)}\;
U(f_{m}\,\overline{f_{m-k_0}}\,f_{m}\,f_{m-k_0}\,;\,
f_{k_0}\,\nu)\;.
\earr \label{snt-13}
\ee
The moments $M_{PQ}$ can be converted into reduced (scale free) cumulants
$k_{PQ}$ that gives information about the shape of the bivariate transition 
strength density. For our purpose the first non-trivial cumulants are the
fourth order cumulants and they are given by,
\be
\barr{c}
k_{40} = \mu_{40} - 3\,, \;\;k_{04} = \mu_{04} - 3\,,\\
k_{31} = \mu_{31} - 3\,\xi\,,\;\;k_{13} = \mu_{13} - 3\,\xi\,,\\
k_{22}= \mu_{22}- 2\;\xi^2 -1 \;; \\
\mu_{PQ} = \l\{\l[\wtM_{20}\r]^{P/2}
\l[\wtM_{02}\r]^{Q/2}\r\}^{-1}\;\wtM_{PQ}\;\;\mbox{and}\;\; 
\wtM_{PQ}=M_{PQ}/M_{00} \;. 
\earr \label{snt-14}
\ee
The $k_{PQ}$, $P+Q=4$ follow from Eqs. (\ref{strn-eq7a}), (\ref{strn-eq7b}),
(\ref{snt-3}), (\ref{snt-7}),(\ref{snt-8}), (\ref{snt-11}), (\ref{snt-12})
and (\ref{snt-13}).
Numerical results for some typical values of $(N,m,k,k_0)$ are shown in
Table 1. These results show that in general $|k_{PQ}| \lazz 0.3$ indicating
that the bivariate strength density will be close to a bivariate Gaussian. For
further confirming this result, we will derive asymptotic results for
$k_{PQ}$. 

\begin{table}[ht]

\caption{Bivariate correlation coefficient ($\xi$) and fourth order bivariate
cumulants $k_{rs}$ with $r \geq s$ and $r+s=4$ for various values of number of
sp states ($N$), number of fermions ($m$), Hamiltonian body rank ($k$) and the
rank ($k_0$) of the particle removal transition operator. Results are obtained 
using the formulas given in Section 3. Note that for the $M_{22}$ that is needed
for  $k_{22}$, we have used Eq. (\ref{snt-13}) with the third term replaced by
the corresponding asymptotic formula given by Eq. (\ref{asymp-snt6}) as a
formula for the reduced matrix elements in Eq. (\ref{snt-13}) is not
available.} 

\begin{center}
\begin{tabular}{cccccccccc}
\\
\\
\hline 
$N$ & $m$ & $k$ & $k_0$ & $\xi$ & $k_{40}$ & $k_{04}$ & $k_{31}$ & 
$k_{13}$ & $k_{22}$ \\
\hline \\
$20$ & $10$ & $2$ & $1$ & $0.82$ & $-0.54$ & $-0.55$ & $-0.44$ & $-0.45$ &
$-0.21$ \\
$30$ & $10$ & $2$ & $1$&  $0.85$ & $-0.48$ & $-0.50$ & $-0.41$ & $-0.43$ &
$-0.26$ \\
$60$ & $10$ & $2$ & $1$ & $0.88$ & $-0.42$ & $-0.46$ & $-0.37$ & $-0.40$ &
$-0.30$ \\
$80$ & $10$ & $2$ & $1$ & $0.88$ & $-0.41$ & $-0.45$ & $-0.36$ & $-0.39$ &
$-0.31$ \\
$50$ & $12$ & $2$ & $1$ & $0.89$ & $-0.38$ & $-0.40$ & $-.034$ & $-0.36$ &
$-0.25$ \\
& $15$ & $2$ & $1$ & $0.91$ & $-0.33$ & $-0.35$ & $-0.30$ & $-0.31$ & $-0.19$ 
\\
& $20$ & $2$ & $1$ & $0.92$ & $-0.29$ & $-0.29$ & $-0.26$ & $-0.27$ & $-0.13$
\\
& $25$ & $2$ & $1$ & $0.92$ & $-0.27$ & $-0.27$ & $-0.25$ & $-0.25$ & $-0.08$ \\
$24$ & $8$ & $2$ & $1$ & $0.82$ & $-0.56$ & $-0.61$ & $-0.46$ & $-0.49$ &
$-0.31$ \\
& & $2$ & $2$ & $0.66$ & $-0.56$ & $-0.67$ & $-0.37$ & $-0.43$ & $-0.22$ \\
$40$ & $15$ & $2$ & $1$ & $0.90$ & $-0.36$ & $-0.37$ & $-0.32$ & $-0.33$ &
$-0.18$ \\
& & $2$ & $2$ & $0.80$ & $-0.36$ & $-0.38$ & $-0.29$ & $-0.31$ & $-0.12$ \\
$60$ & $20$ & $2$ & $1$ & $0.93$ & $-0.27$ & $-0.27$ & $-0.25$ & $-0.25$ &
$-0.14$ \\
& & $3$ & $1$ & $0.89$ & $-0.51$ & $-0.53$ & $-0.46$ & $-0.47$ & $-0.30$ \\
& & $3$ & $2$ & $0.79$ & $-0.51$ & $-0.54$ & $-0.40$ & $-0.43$ & $-0.22$ \\
\hline
\end{tabular}
\end{center}
\label{tab1}
\end{table}

\section{Asymptotic formulas for bivariate moments and approach to bivariate 
Gaussian form}

Here we will consider the asymptotic limit defined by $N \rightarrow \infty$
with $m$, $k$ and $k_0$ fixed and $k$, $k_0 << m$. Note that in the dilute limit
(or true asymptotic limit) we  also have $m \rightarrow \infty$ and  $m/N
\rightarrow 0$ with $k$ and $k_0$ fixed. Using the formulas given in Sections 2
and 3 first we can show that in the asymptotic limit (asymp),
\be
\barr{l}
{\binom{N}{m}}^{-1}\binom{N-k_0}{m-k_0}\;Z_{11}(N,m,k_0,k,k) =
\l[\binom{N}{k_0}\,d(N:k)
\,\Lambda^k(N,m,m-k)\,\Lambda^k(N,m-k_0,m-k_0-k)\r]^{1/2} \\  
\\
\times \;{\binom{N}{m}}^{-1}\binom{N-k_0}{m-k_0}\;
\l| U(f_{m}, \overline{f_{m-k_0}}, f_{m}, f_{m-k_0}\,;\,f_{k_0}, k)\r|\;
\stackrel{\mbox{asymp}}{\longrightarrow}  \binom{m}{k}\,\binom{N}{k}\,
\binom{m-k}{k_0}\;,\\
\\
\overline{\lan [H(k)]^2\ran^m} = \Lambda^0(N,m,k) \stackrel{\mbox{asymp}}{
\longrightarrow}  \binom{m}{k}\,\binom{N}{k}\;,\\
\\
{\binom{N}{m}}^{-1}\,\Lambda^k(N,m,m-k)\,\Lambda^k(N,m,k)\,d(N:k) 
\stackrel{\mbox{asymp}}{\longrightarrow} \binom{m}{k}\,\binom{m-k}{k}
\,{\binom{N}{k}}^2\;.
\earr \label{asymp-snt1}
\ee
Starting with $\xi$, it should be clear that in the asymptotic limit only the 
term with $\nu=k$ in Eq. (\ref{snt-8}) will survive. Then, applying the first 
two relations in Eq. (\ref{asymp-snt1}) will give 
\be
\barr{l}
\xi(m) = \dis\frac{M_{11}(m)}{M_{00}(m)\,\l[\wtM_{20}(m)
\wtM_{02}(m)\r]^{1/2}} \\
\\
\stackrel{\mbox{asymp}}{\longrightarrow} \dis\frac{\binom{m-k}{k_0}\; 
{\binom{m}{k}}^{1/2}}{\binom{m}{k_0}\;{\binom{m-k_0}{k}}^{1/2}}\;. 
\earr \label{asymp-snt2}
\ee
Similarly, for $k_{40}$ and $k_{04}$ only the terms with $\nu=k$ in Eq. 
(\ref{strn-eq7b}) will survive and then applying the second and third relations 
in Eq. (\ref{asymp-snt1}) will give,
\be
\barr{l}
k_{40}(m) = \dis\frac{\wtM_{40}(m)}{\l[\wtM_{20}(m)\r]^2} \;-3
\stackrel{\mbox{asymp}}{\longrightarrow} \dis\frac{\binom{m-k}{k}}{
\binom{m}{k}} -1\;,\\
\\
k_{04}(m) = \dis\frac{\wtM_{04}(m)}{\l[\wtM_{02}(m)\r]^2} \;-3
\stackrel{\mbox{asymp}}{\longrightarrow} \dis\frac{\binom{m-k_0-k}{k}}{
\binom{m-k_0}{k}} -1\;.
\earr \label{asymp-snt3}
\ee
For $M_{31}$, the first term in Eq. (\ref{snt-11}) is trivial and in the sum in
the second term only the $\nu=k$ term will survive in the asymptotic limit. Now,
applying Eqs. (\ref{asymp-snt1}) and (\ref{reduced}) will give the result for
$k_{31}(m)$,
\be
k_{31}(m) \stackrel{\mbox{asymp}}{\longrightarrow} \dis\frac{\binom{m-k}{k} 
\binom{m-k}{k_0}}{\binom{m}{k_0}\;\dis\sqrt{\binom{m}{k}\,\binom{m-k_0}{k}}} 
-\xi(m)= \xi(m)\,k_{40}(m)\;.
\label{asymp-snt4}
\ee
Similarly $k_{13}(m)$ is given by,
\be
k_{31}(m) \stackrel{\mbox{asymp}}{\longrightarrow} \dis\frac{\binom{m-k_0-k}{k}
\binom{m-k}{k_0}\;{\binom{m}{k}}^{1/2}}{\binom{m}{k_0}\;{\binom{m-k_0}{k}
}^{3/2}} -\xi(m) = \xi(m)\,k_{04}(m)\;.
\label{asymp-snt5}
\ee
Finally, in $M_{22}$ only the third term in Eq. (\ref{snt-13}) is complicated. 
This is simplified using its relation, valid in the asymptotic limit, 
to $\xi(m)$ as described in \cite{group30}. Following this we have for $k_{22}$,
\be
\barr{l}
k_{22}(m) \stackrel{\mbox{asymp}}{\longrightarrow} -2\,[\xi(m)]^2 +
\dis\frac{\binom{m}{k}\,{\binom{m-k}{k_0}}^2}{\binom{m-k_0}{k}\,
{\binom{m}{k_0}}^2}
+\dis\frac{\binom{m-2k}{k_0}\,\binom{m-k}{k}}{\binom{m}{k_0}\,\binom{m-k_0}{k}}
\\
\\
\approx -2\,[\xi(m)]^2 + \dis\frac{\binom{m-2k}{k_0}}{\binom{m}{k_0}\,
\binom{m-k_0}{k}}\l[\binom{m}{k} + \binom{m-k}{k}\r]\;.
\earr \label{asymp-snt6}
\ee
In the dilute limit with $m \rightarrow \infty$ and $m/N \rightarrow 0$ and
expanding the binomials in Eqs. (\ref{asymp-snt2}) to (\ref{asymp-snt6}), 
it is seen that to order $1/m$ the cumulants $k_{rs}$, $r+s=4$ will be $-k^2/m$
(independent of $k_0$) and the correlation coefficient $\xi(m) \rightarrow 1- 
(k k_0)/2m$. Thus, the cumulants will tend to zero giving bivariate Gaussian
form. However, as $\xi \rightarrow 1$ as $m \rightarrow \infty$, in practice it
is necessary to add the $k_{rs}$, $r+s=4$ corrections to the bivariate Gaussian.
This is same as the result seen for $t$-body transition operators before in
\cite{group30,FKPT}.

\section{Extensions to particle addition operators and beta decay type 
operators} 

Firstly, particle addition operator is $\co^+=\sum_\alpha V_\alpha \, 
A^\dagger_{\alpha}(k_0)$ and acting on a $m$-particle state it will generate
$m+k_0$ particle states. It is easy to see that the formulation in Section 3
will apply directly to $\co^+$ operator by appropriately changing everywhere
$m-k_0$ by $m+k_0$ giving formulas for $\xi$ and $M_{rs}$, $r+s=4$. Explicit
formulas are not given here due to lack of space.

Now we will turn to beta decay type operator and for this
consider a system with $m_1$ fermions in $N_1$ sp states and $m_2$
fermions in $N_2$ sp states with $H$ preserving $(m_1,m_2)$. Then, the $H$ 
operator, assumed to be $k$-body, is given by,
\be
\barr{l}
H(k) = \dis\sum_{i+j=k} \dis\sum_{\alpha , \beta \in i} \dis\sum_{a,b \in j}\;
V_{\alpha a:\beta b}(i,j)\;
A^\dagger_{\alpha}(i)\, A_{\beta}(i)\, A^\dagger_a(j)\,A_b(j)\;,\\
\\
V_{\alpha a:\beta b}(i,j) = \lan i, \alpha : j, a \mid H \mid i, \beta : 
j,b\ran\;.
\earr \label{strn-eq14}
\ee
Here we are using Greek labels $\alpha, \beta, \ldots$ to denote the many
particle states generated by fermions occupying the orbit with $N_1$ sp states
and the Roman labels $a,b,\ldots$ for the many particle states generated by the
fermions occupying the orbit with $N_2$ sp states.  For each $(i,j)$ pair with
$i+j=k$, we have a matrix $V(i,j)$ in the $k$-particle space and we assume that
the $V(i,j)$ matrices are represented by independent GUE's with 
their matrix elements being zero centered with variance,
\be
\overline{V_{\alpha a : \beta b}(i,j)\;V_{\alpha^\pr a^\pr : \beta^\pr b^\pr}
(i^\pr ,j^\pr)} 
=V^2_H(i,j)\;\delta_{i i^\pr} \delta_{j j^\pr} \delta_{\alpha \beta^\pr} 
\delta_{a b^\pr} \delta_{\beta \alpha^\pr} \delta_{b a^\pr} \;.
\label{strn-eq15}
\ee
It is important to note that the embedding algebra for the EGUE generated by 
the action of the $H(k)$ operator on $\l.\l|m_1, v_\alpha : m_2, v_a\r.\ran$ 
states is the direct sum algebra  $U(N_1) \oplus U(N_2)$. Thus we have 
EGUE($k$)-$[U(N_1) \oplus U(N_2)]$ ensemble. A beta decay type transition 
operator is given by
\be
\co = \dis\sum_{\alpha , a} O_{\alpha a} A^\dagger_{\alpha}(k_0)\,
A_a(k_0)\;;\;\;\;O_{\alpha a} =
\lan k_0,\alpha \mid \co \mid k_0, a\ran\;.
\label{strn-eq16}
\ee
Note that for beta decay $k_0=1$ and for double beta decay $k_0=2$ in Eq.
(\ref{strn-eq16}). To proceed further, we assume a GUE representation for 
the $\co$ matrix in the defining space
giving $\overline{O^\dagger_{\alpha , a} O_{\beta , b}}=
V^2_\co\;\delta_{\alpha \beta}  \delta_{a b}$.  Note that in general the $\co$
matrix is a rectangular matrix. Now,  the ensemble averaged bivariate moments of
the transition strength density are $M_{PQ}(m_1 , m_2)=\overline{\lan \cod H^Q
\co H^P\ran^{(m_1 , m_2)}}$. Note that $\co$ takes $(m_1 , m_2)$ to $(m_1+k_0 ,
m_2-k_0)$. Formulas for the moments will follow by applying the formulation for
particle removal operator (given in Section 3) in $m_2$ space and for particle
addition operator in $m_1$ space with appropriate summations over different parts
of $H$. Using this, formulas are derived for $M_{PQ}$ with $P+Q=2$ and $4$ and
there will be reported elsewhere. 

\section{Conclusions and future outlook}

In this paper, we have presented exact (finite $N$) results for the moments of
the transition strength densities generated by particle removal operators, using
$U(N)$ Wigner-Racah algebra for EGUE random matrix ensembles for spinless
fermion systems.  In particular, formulas for the moments up to fourth order are
derived in detail for the Hamiltonian a EGUE($k)$ and a $k_0$ number of
particles removal transition operator with its structure coefficients in the
defining spaces [$V_{\alpha_0}$ in Eq. (\ref{snt-1})] assumed to be  independent
Gaussian variables. Numerical results on one hand and the asymptotic results
derived from the exact results on the other, showed that the fourth order
cumulants approach zero in the dilute limit implying that the strength densities
approach bivariate Gaussian form. As discussed briefly in Section 5, the
formulation given in  Sections 3 extends to transition operators  that are
particle  addition operators and also to beta decay and neutrinoless double beta
decay type operators. Results of the present work, the results reported in the
Ghent meeting in June 2014 \cite{group30} where the transition operator is a
$t$-body operator represented by EGUE($t$) and the results (briefly reported in
\cite{Ma-arx} and to be  reported in detail in a long paper in preparation) for
beta decay and double beta decay type operators establish clearly that the form
for the bivariate transition strength densities for isolated finite fermion
systems will be {\it generically a bivariate Gaussian}. Therefore, the bivariate
Gaussian form with some corrections can be used in practical applications in
calculating transition strengths in complex systems just as the corresponding
results for level densities are being applied for calculating nuclear level
densities by the Michigan group \cite{Senkov1,Senkov2}. Finally, further
extensions of the present work to EGUE with $U(\Omega) \times SU(r)$ embedding
discussed in \cite{Ma-12} and also to boson systems will be important and they
will have applications in mesoscopic systems [here $r=2$ with fermions is
important] and Bose gases [here $r=1$ and $r=3$ with bosons will be important].

\ack
Thanks are due to Manan Vyas for many useful discussions.

\end{document}